# THE DILEMMA OF BOSE SOLIDS: IS HE SUPERSOLID?

## ABSTRACT

Nearly a decade ago the old controversy about possible superfluid flow in the ground state of solid He4 was revived by the apparent experimental observation of such superflow. Although the experimentalists have recently retracted, very publicly, SOME of the observations on which such a claim was based, other confirming observations of which there is no reason for doubt remain on the record. Meanwhile theoretical arguments bolstered by some experimental evidence strongly favor the existence of supersolidity in the Bose-Hubbard model, and these arguments would seem to extend to solid He. The true situation thus is apparently extraordinarily opaque. The situation is complicated by the fact that all accurate simulation studies use the uniform sign hypothesis which confines them to the phase-coherent state, which is or could be, in principle, supersolid, so that no accurate simulations of the true classical solid exist. There is great confusion as to the fundamental nature of the ground state wave-function for a bose quantum solid, and we suggest that until that question is cleared up none of these dilemmas will be resolved.

## INTRODUCTION

Nearly a decade ago[1] the old conjecture that the Bose solid He-4 would exhibit superflow was revived due to a series of experiments with torsional oscillators (TO) at 10-200 mdeg K which were confirmed worldwide, though with somewhat inconsistent results. Disturbingly, it was found that the phenomenon was extraordinarily sensitive to small impurity concentrations of He-3. 4 years later, it was observed that with apparently very similar temperature and impurity dependence the shear stiffness of He-4 rises sharply at low T.

Simulation studies of high accuracy on pure He-4 have, on the whole, failed to find any evidence of these phenomena, and a view has become popular that the whole suite of effects is caused by dislocation pinning and unpinning and that supersolidity has no theoretical validity. This view was reinforced recently by the very public retraction[2] of some—but not all[3]-- of the TO results by the originator, Moses Chan, and very careful and definitive studies on single crystal elasticity reinforce this view, though themselves presenting some interesting dilemmas.[4]

A NOTE ABOUT SIMULATIONS
In 2004, at the time of the original observations, there was no definitive theoretical treatment of Bose solids available other than simulations. A feature of the simulations which has not been given enough notice is that, in a sense, they beg the question of superfluidity: the success of the simulations in reproducing the physical properties of solid He is without exception based on taking advantage of the absence of the "sign problem" for Bose systems. The assumption is made that the wave function does not have any zeroes, and that the wave function is real and positive everywhere. Thus every simulation is of a sample which is totally free of vortices, which are zeroes of the complex amplitude of the Bose field, while a true ""classical" solid is by definition a condensate of vortices: no accurate simulation has thus ever been carried out of an actual physical sample of solid He at, say, .1 to 1 degree K, and we have no comparison, via simulations, of the two states of matter, solid helium without vortices and with rigid phase restrictions, as compared to helium without rigid phase in which the bosons can be confined to their separate sites.

# HEURISTIC HAMILTONIAN

The present author proposed a heuristic Hamiltonian[5] for supercurrents[6] but until recently[7] had not given it a particularly sound foundation. This Hamiltonian describes the supercurrents as the result of a dependence of the energy on a phase variable φ(r) which is the average phase of the local Bose field ψ(r). In a lattice, the simplest form for such a Hamiltonian would be the "x-y" model,

$$H_{xy} = -\sum_{i,j} J_{i-j} \cos(\varphi_i - \varphi_j) \text{ and } j = \nabla H_x \text{ is the particle current. [1]}$$

The current will be divergenceless $\nabla \cdot j = \nabla^2 \varphi = 0$

with reference to the site lattice, because the frequency scale of J is much smaller than the Debye frequency, so that atoms cannot accumulate and any motions on a slow time scale have to be incompressible.

If [1] is a component of the Hamiltonian, at T=0 φ will be uniform, $\varphi_i = \varphi_j$. Rotation will lead to a vector potential which causes a supercurrent to flow, as described in my earlier paper, so this is a superfluid in that sense. At some T of order J the system will undergo a thermal phase transition, in 2D of BKT type or in 3D a 3D x-y transition. But the dynamics is not trivial because the heuristic Hamiltonian [1] refers to phases on the sites; and for instance a static shear of the site lattice does not cause a vector potential in the phase field. As is the case with superconductors, the phase order parameter is a topological one, not a locally determined object: it can only be affected by overall rotation or a change in boundary conditions. The reason why currents flow with ease into and out of superconductors is the existence of Andreev scattering which converts pairs into normal current, and there is nothing like that here. In a pure sample only the equivalent of diamagnetic currents can flow. (but see below)

BOSE-HUBBARD MODEL
In the past few years another avenue toward Bose solids has become available: the Bose-Hubbard (B-H) model, which can be accurately modeled by cold atoms in an optical lattice.[8] A very straightforward argument[9] can be constructed that the Bose-Hubbard model in its ground state always has at least a small superfluid density if D≥2 even when the interaction parameter U/t is quite large, and the experimental observations on cold atoms, and simulations, favor this conclusion as we will explain below, although they have not yet been adequately analyzed.

The difference between the Bose-Hubbard model and a true Bose solid is that the former has a predetermined periodic lattice while the lattice of the solid is formed self-consistently. The effect is to eliminate phonons as excitations in the B-H model. But as we said above, phonons do not couple very effectively to these divergenceless currents. A second, and mathematically very simplifying, difference is that the Hilbert space of wave-functions in the B-H model is restricted to be only N-dimensional, where N is the number of lattice sites: one state and one boson wave function per site. Nonetheless this system has enough freedom that it can describe a superfluid, a true solid, and, as we will see, a supersolid.

The Bose-Hubbard model consists of the Hamiltonian

$$H = \sum_{i,j} t_{ij} b^*_i b_j + \sum_i U n_i(n_i - 1) - \mu \sum_i n_i \quad [2]$$

It is well understood that for large values of U/t and low T the only stable phases are those with integer <n,>, and μ=U(<n>-1). The simple case n=1, μ≅0 exhibits all the interesting physics. The obvious trial solution for the ground state is

$$\Psi_0 = \prod_i b_i^* |vacuum\rangle \qquad [3]$$

and the single-particle elementary excitation energies are determined by the equations of motion of holes and particles:

$$(E_i - E_c) b_i^* \Psi_0 = [H, b_i^*] \Psi_0 = \left[ \sum_j t_{ij} b_j^* + U b_i^* \right] \Psi_0 \qquad [4]$$

for particles; and

$$(E_0 - E_i) b_i \Psi_0 = \sum_j t_{ij} b_j \Psi_0 \qquad [5]$$

for holes.

These two equations, if taken literally and solved straightforwardly, simply give us two bands of running-wave solutions with an intrinsic gap of U. But this is not the correct approach to the Hartree-Fock conceptual structure in the Bose solid case. The Hartree-Fock concept, of seeking a product trial function, is straightforward if one is looking variationally for a product of extended wave functions, and in the Fermion case thanks to Wannier's theorem and the exclusion principle this is always possible. But in the Bose case the correct trial function is of the form [3] of a product of *local* functions. Unlike the Fermion case, there is no equivalence to a product of Bloch extended functions. These local functions are not orthonormal and are not equivalent to a product of running Bloch waves; each satisfies a *different* wave equation, as I pointed out in[10].

The Bose-Hubbard model starts with the assumption that the local functions are an orthonormal set; it is equivalent to

defining a set of orthogonal Wannier functions φ(r-r$_i$) for the lattice, and defining

$$b_i = \int \varphi(r - r_i) d^D r \quad [6]$$

which obey canonical commutation rules; and assuming that the set of φ's is sufficiently complete to describe most low-energy behavior.

The trial function [3], however, is clearly not the ground state, because the Hamiltonian contains the matrix elements t$_{ij}$ which connect to states with doubly-occupied sites. Therefore the above solution is only metastable. Following Kohn[11], we attempt to eliminate the matrix elements between low-energy and high-energy states perturbatively in succeeding orders of t/U, but in this case we cannot use a unitary matrix e$^{iS}$ for the canonical transformation.; we must make a linear, nonunitary transformation of the b$_i$'s into a *nonorthogonal* set b$_i$':

$$b_i' = (b_i + \sum_{j \neq i} S_{ij} b_j) / \|1 + S\| \quad [7]$$

and || || is the permanent. Our trial function now is

$$\Psi_0' = \prod_j b_j'^* |vacuum\rangle \quad [8]$$

The overlap coefficients are to be determined perturbatively in powers of t/U. A possible procedure is to ask for an equation of motion for a hole, in the potential which exists *after* the particle has been removed. (This is similar to the procedure in ref 10.) We replace [5] by

$$(E_o - E) b_i' \Psi_0' = [\sum_j t_{ij} b_j' - \sum_{j \neq i} U b_j] \Psi_0' \quad [9]$$

In other words, the same equation as [4] except that the particle at site i has been removed, since it cannot experience its own repulsive potential when it is part of the ground state.

A procedure essentially equivalent to this one was carried out to a high degree of accuracy as a series in t/U by H Monien and N Elstner[12] some years ago, essentially equivalent to iterating equations 4, 5 and 9 perturbatively. Their work amounts to a proof that the iterative procedure based on localized orbitals [7] converges as a series in t/U up to a critical value, which is the value at which the "Mott solid" no longer exists.

How do we demonstrate the existence of superflow in this wave function? The criterion we use was proposed by Kohn, who used it to demonstrate the *non*existence of flow in a Fermionic Mott insulator: whether the energy is affected by a change in boundary conditions, which is equivalent to applying a uniform vector potential A to the system, which is equivalent to rotating a toroidal sample. As is well-known, dE/dA=J.

For a conventional Fermion solid or insulator, we know that the bands are either full or empty, the crystal momentum states are equally occupied and a uniform shift by A cannot change the kinetic energy (see figure). However, if the Wannier functions are non-orthogonal, the crystal momentum distribution is not uniform but weighted to lower values (the momentum distribution P(k) varies as $-S\cos k\, r_{ij}$) and the net kinetic energy, and with it the potential energy, changes when k->k+A.[13]

The quantity p(k) is directly measured in cold atom experiments, and in fact in the higher range of t/U the measurements seem to exhibit a diffuse peak just above where the sharp condensate line disappears. (see figure) The measurement is complicated a bit by the sample inhomogeneity, but the existence of a peak is supported by the fact that it appears in Monte Carlo simulations of the Bose-Hubbard model as well.  The reasoning above as to why such a peak implies supersolidity is simple and rigorous and seems to give strong, if indirect, support to the existence of supersolidity in this case.

ELASTICITY

The parameters $J_{ij}$ are functions of the interparticle distance $r_{ij}$, of course, and therefore when the phase is ordered its elastic constants are not the same as those of the randomly phased conventional solid.  This obvious fact seems to have been ignored by all who have discussed the controversy, and it has been universally assumed that the elasticity results contradict supersolidity.  Qualitatively, the phase-ordered solid should be stiffer than the conventional one, since the motivation for phase-ordering is to increase the binding energy; and therefore the results of Beamish are more  confirmation than otherwise.  The problem of course is explain the magnitude of the effect, and as I said above none of the simulations are of any value in this.  The elastic constants depend on second derivatives of J with respect to r and thus may be extremely sensitive.  Also, hexagonal He has a remarkably soft  $c_{44}$, so r-dependence of the J's will have a non-negligible effect.

HE3

Finally, let me bring up another effect which has not  been discussed at all in the theoretical literature, the effect of supersolidity on He3 "impurities".  It is easily understood that if the solid is a 'condensate of vortices" so that the hopping matrix elements are phase-averaged out , <u>both </u>of the heliums are essentially localized quantum-mechanically.  But in the phase-ordered state it becomes possible for He3 to undergo quantum diffusion.  That is, there is a matrix element connecting the state with He3 localized near site i to that with it on site j.  In the simplest Bose-Hubbard version of this situation, we have a hopping integral for He3 of t', while He4 is t;  the effective bandwidth for He3 will then be tt'/U, taking the need for backflow into account.  One would estimate that t' will be considerably larger than t, so that the He3 bandwidth will be bigger than J.  Unfortunately it is not obvious how to take advantage of this to explain the magnitude of the effects of He3.  What we can say is that He3 will become a mobile entity when He3 turns supersolid.

There is a way to test experimentally whether the scenario suggested by the above considerations is correct.  I have repeatedly suggested that high-sensitivity NMR studies on the He3 impurity be undertaken, and it appears that this change in diffusion behavior may be a reliable test for supersolidity.

There is a popular alternative scenario centered around dislocation pinning by He3 impurities which has been advanced to explain the elasticity results, in particular the very detailed studies of large single crystals carried out by Balibar, Beamish et al., which show that it is only the $c_{44}$ modulus which seems to have the strong temperature dependence (see figure, from reference 5).  This is the modulus which represents shear of the hexagonal planes relative to each other, and would be affected by arrays of dislocations (or half-dislocations) lying in these planes.  The measurements are very striking and their details suggest strongly that dislocations are part of the story.  But recently[14] they have discovered that there is a transition when the pinning centers themselves become mobile.

CONCLUSIONS AND THOUGHTS

Many of the difficulties in understanding the data come from looking for the wrong thing. ODLRO , for instance, is not a property of the supersolid state; it isn't even a property of superconductors.[15]  I was among those captured, early on, by the idea that physical vacancies were necessary, and they are not. The relevant phase comes in via the interference between the local bosons of the ground state,  which occurs when they are not orthogonal and hence do not commute.  In every way, coherence  makes the lattice *more* stable: for instance, the nonorthogonal local functions are more localized and smaller than orthogonal ones, and may even interchange less often. The improvement in binding energy may not be negligible, though it is a mystery that it shows up so little in the specific heat—but are we sure that the phase system is in equilibrium? The Chan specific heat peak[16] takes on real importance.  I don't think any  of the dislocation array theories account for enough entropy to explain it.

Impurities make superconductors work better;  why not supersolids?  Are they—and dislocations—pinning vortices and not dislocations, or both?  Or are they, as I suspect, necessary to achieve equilibrium?

In general, I would like to encourage people to look at the real, theoretically sound possibilities; for instance cold atoms need to be carried to much colder temperatures to confirm NCRI, which is firmly and quantitatively predicted by theory in that case.

ACKNOWLEDGEMENTS

I must acknowledge extensive and fruitful  discussions with Bryan Clarke, Shivaji Sondhi, David Huse, Nandini Trivedi and Minoru Kubota.

FIGURE CAPTIONS
1) Schematic of effect of rotation on momentum distribution

2) Momentum distribution in cold atom experiment

3) Temperature dependence of $c_{44}$ of ultrapure He crystal

Superfluid: One momentum state occupied
Insulator: All momenta equally occupied
Supersolid: Low momenta preferentially occupied

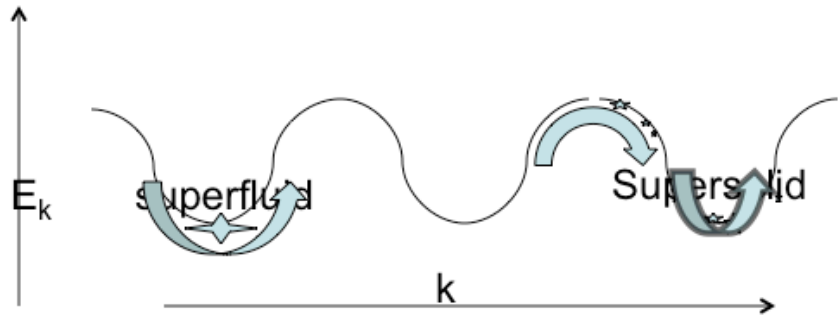

But the supersolid has no ODLRO: the density matrix is short-range

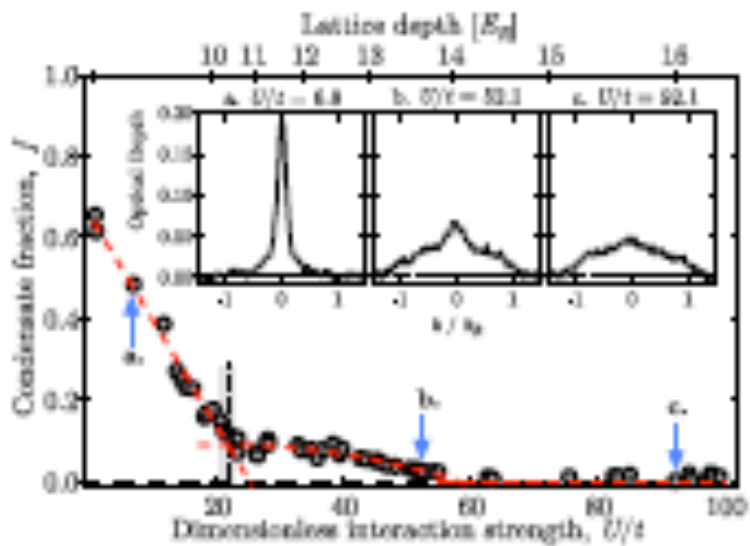

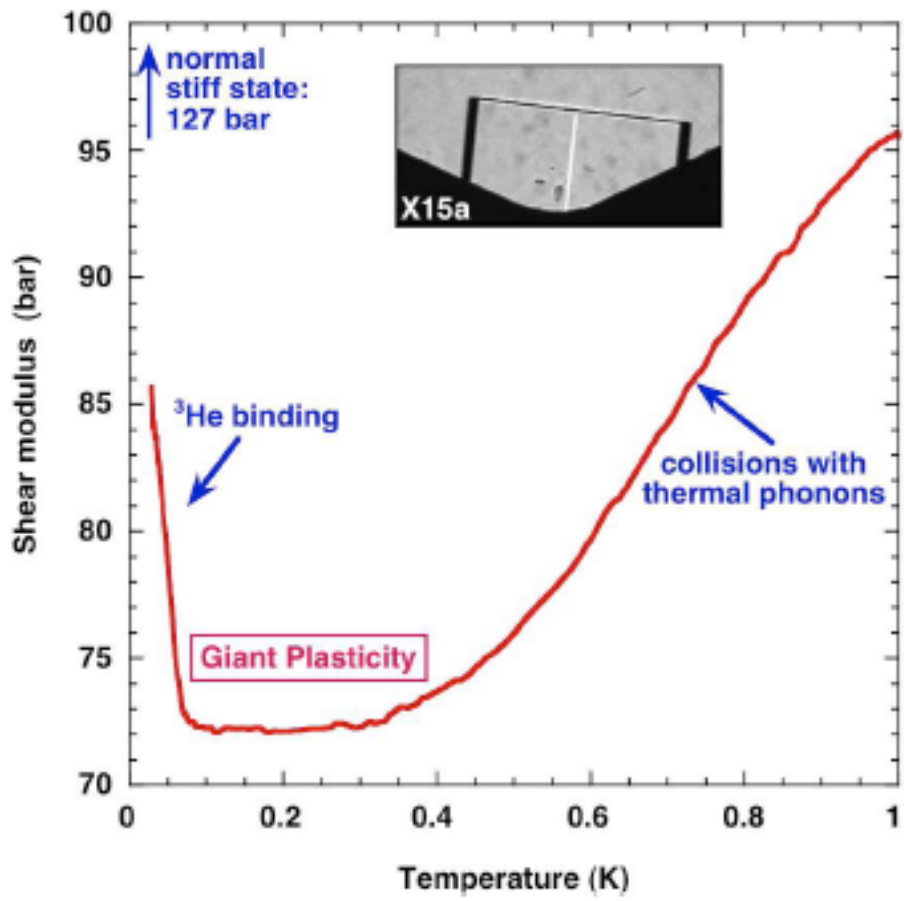